\documentclass[doublecol]{epl2}
\usepackage{epsf,amssymb,amsfonts,amsmath,graphics}
\usepackage{verbatim}

\title{Modeling the mobility with memory}
\shorttitle{Mobility with memory}
\author{Jeehye Choi \and Jang-Il Sohn \and K.-I.~Goh\thanks{\email{kgoh@korea.ac.kr}} \and I.-M.~Kim} 
\shortauthor{J. Choi \etal}
\institute{Department of Physics, Korea University, Seoul 136-713, Korea}

\abstract{
We study a random walk model in which the jumping probability to a site is dependent on the number of previous visits to the site, as a model of the mobility with memory. To this end we introduce two parameters called the memory parameter $\alpha$ and the impulse parameter $p$. From extensive numerical simulations, we found that various limited mobility patterns such as sub-diffusion, trapping, and logarithmic diffusion could be observed. By the memory, a long-ranged directional anti-correlation kinetically-induces anomalous sub-diffusive and trapping behaviors, and transition between them. With random jumps by the impulse parameter, a trapped walker can escape from the trap very slowly, resulting in an ultraslow logarithmic diffusive behavior. Our results suggest that the memory of walker's has-beens can be one mechanism explaining many of empirical characteristics of the mobility of animated objects. 
}

\begin{document}
\maketitle

\section{Introduction}
Since Einstein's study on the motions of small particles suspended in liquids \cite{einstein}, diffusion and random walk have been a paradigm in describing and modeling the mobility involving seemingly irregular motions exhibited by both material and animated objects \cite{avraham}. A multitude of variations of the random walk model with additional details  have been widely studied \cite{hughes,bouchaud}, covering diverse physical and biological domains ranging from charge transport in disordered conductors to foraging patterns of animals and fishes.

From the random-walk perspective, even the movement of a human individual may look irregular and random. In contrast to material objects, however, humans are conscious beings and driven by motivations by their will and purposes modulated by external cause, so their mobility patterns cannot be completely random.
Indeed, recent advances in digital technology provided us with a plenty of empirical data for human mobility \cite{brockmann,gonzalez}, which revealed a number of key empirical characteristics of human mobility patterns, three of which we highlight here:
First, even though the individual trajectory bears some degree of randomness, 
most part of it is highly predictable, being embedded in a well-defined region in space with the radius of gyration of the trajectory growing logarithmically in time \cite{gonzalez}.
Second, the mobile object (a human individual in this case) spends 
disproportionate times in different locations \cite{song}.
Third, at the population level, there is high degree of heterogeneity across individual's degree of mobility \cite{gonzalez}.

In this work, our key premise is that at the heart of these human mobility patterns does underlie the role of {\em memory}: Conscious beings do have memory of where he or she has been and tend to revisit some of the have-beens, his or her favorite spots. By incorporating such a memory effect in the dynamic rule of a random walk model, we investigate the consequential role of memory in the mobility patterns.

We will demonstrate that the memory induces a long-ranged directional anti-correlation, which kinetically slows down the walker's mobility, resulting in sub-diffusion and trapping, and transition between them. Logarithmic diffusion can occur by introducing the impulsive random jumps. Therefore, many of characteristics of limited mobility can be explained by simple rules based on memory.

\section{Model}
Our model is defined on a hypercubic lattice in $d$-dimensions.
Starting from the origin, the walker takes a random walk with jumping probability depending on the number of previous visitations to the target sites. Specifically, we denote $n_i(t)$ the number of visitations to the site $i$ upto $t$ steps by a random walker starting from the origin at time $t=0$. The jumping probability $w_{j,i}(t)$ from the site $i$ to site $j$ at time $t$ is given by
\begin{equation}
w_{j,i}(t) = \left \{ \begin{array}{ll}
[n_j(t)+1]^{\alpha}/Z & \textrm{with probability $1-p$}\\
1/(2d) & \textrm{with probability $p$},
\end{array}
\right.
\end{equation}
with the normalization $Z={\sum_j}^{\prime}[n_j(t)+1]^{\alpha}$ by the restricted sum over nearest neighbors $j$ of the site $i$, and $d$ the spatial dimension.
Thus the jumping probability changes dynamically, depending on the history of the random walk.

The two parameters $\alpha$ and $p$ characterize our model. $\alpha$ determines the degree of memory-dependent revisitation tendency. So we call it the memory parameter. The higher $\alpha$ is, the walker tends to go back to sites previously visited more often. The parameter $p$ is introduced to implement occasional purely random jumps, which may occur due to, e.g., impulsive decisions of a mobile walker. So we call $p$ the impulse parameter. With $p=0$, the walker always performs the memory-dependent move. With $p=1$, the model is reduced to simple symmetric random walk. In Fig.~1, typical trajectories from our model in two dimensions with different combinations of $(\alpha,p)$-parameters are shown.

\begin{figure}
\centering
\includegraphics[width=.9\linewidth]{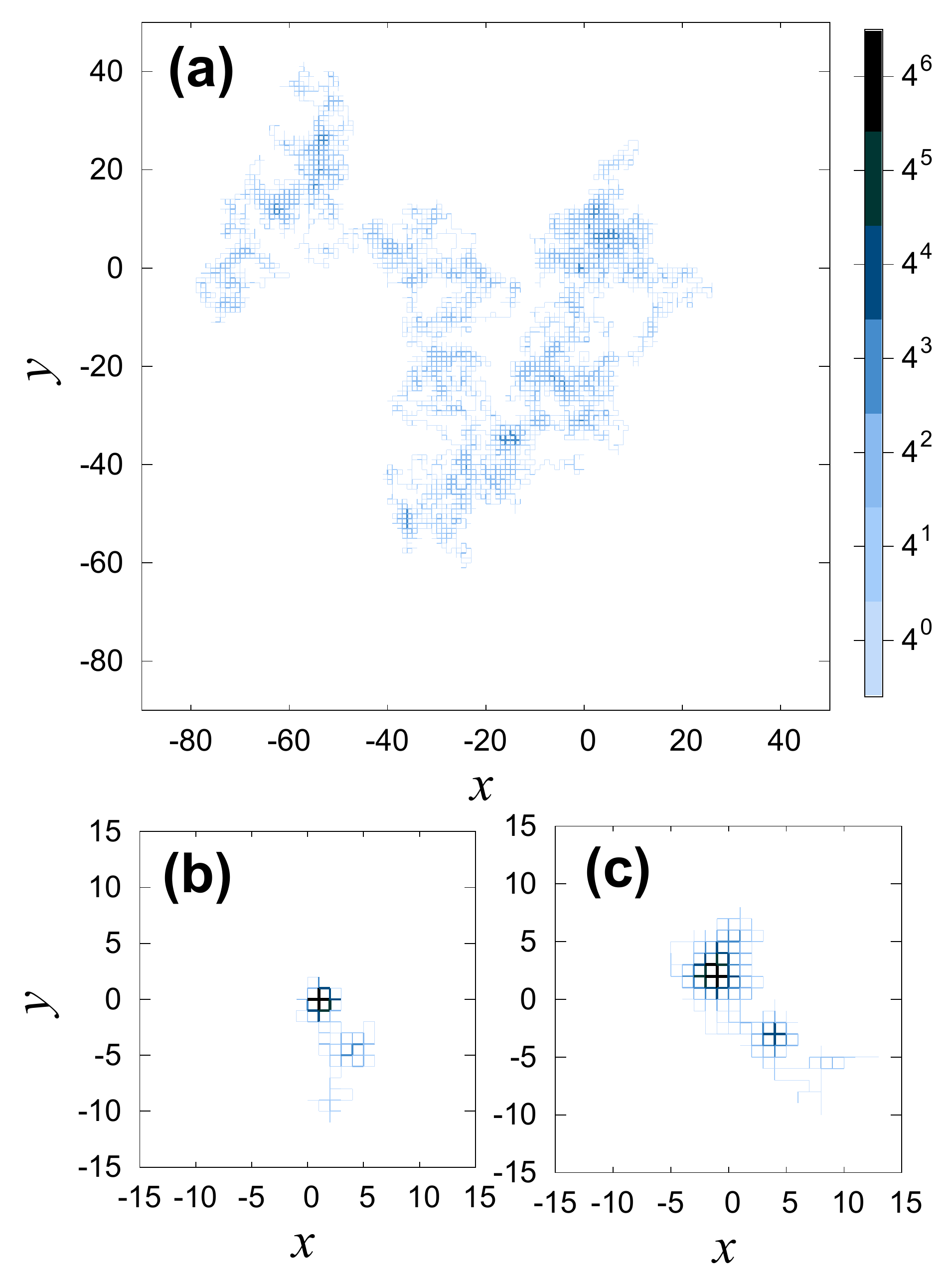}
\caption{Typical trajectories from the model, corresponding to the parameter combinations $(\alpha,p)=(0.4,0)$ (a), $(0.8,0)$ (b), and $(1,0.2)$ (c), each representing the sub-diffusive, trapped, logarithmic diffusive behaviors, respectively. Trajectories upto $t=2^{15}$ in two dimensions are shown, colorcoded by the visitation numbers (see the color-bar). Note the difference in the ranges of view field, as well as the range of variations in visitation numbers in the three cases.
}
\end{figure}

Random walk models in similar vein to ours had been studied in relation to self-interacting random walks \cite{domb,stanley1,duxbury,sapozhnikov,ordemann}.
For example, Duxbury and de Queiroz \cite{duxbury} assigned to each (simple) random walk trajectory the ``energy'' in the form of $\exp\left(-g \sum_i{n_i}^{\alpha}\right)$, where $n_i$ is the visitation numbers of the site $i$ as in our model.
When $g>0$ and $\alpha\ge 0$, constrained trajectories are favored (note that even $\alpha=0$ case is not reduced to simple random walk in this model). Depending on $g$ and $\alpha$, different behaviors occur, such as anomalous sub-diffusion, trapping at origin, or self-avoiding walks \cite{duxbury}.
Although apparently similar, this class of models are equilibrium-type models, whereas our model is a generative model with explicit dynamic rules. 

More recently, human mobility models directly motivated by data from digital records of human travels and mobility have been proposed \cite{hufnagel,colizza,song2}. For example, Song {\it et al.} derived two rules of human mobility from the mobile phone usage data \cite{song2}. They introduced a human mobility model incorporating those rules, which they termed the preferential return and the exploration, similar to our memory and impulse parameters. With such data-driven modeling, they explained many characterisics of empirical human mobility patterns, e.g., the ultraslow logarithmic growth of the radius of gyration of individual's trajectory in time.
This model and ours share key properties in dynamic rules, but the precise implementation is somewhat different. First, the model of Ref.~\cite{song2} is a spatial network model, whereas ours is a lattice model. Second, the parameters in our model cover a broader range, therefore we can address the generic role of these parameters, which might be relevant for novel mobility patterns associated with memory effect, other than those identified thus far.

\section{Role of memory parameter}
The memory parameter $\alpha$ determines the degree of memory-dependent revisitation tendency. We are primarily interested in the cases of positive $\alpha$, for which the walker tends to revisit the sites previously visited many times, resulting in a constrained trajectory [see Fig.~1(a) for example] compared to simple symmetric random walks, similarly to interacting random walks \cite{stanley1,duxbury}. When $\alpha=0$, the model becomes memoryless and is reduced to simple symmetric random walk. When $\alpha$ is negative, the walker disfavors revisitation, which would lead to a stretched trajectory, similarly to weakly self-avoiding walks \cite{domb}.
In the limit $\alpha\to-\infty$, the model becomes the self-avoiding walk.
In the opposite limit $\alpha\to\infty$, which is of more interest to us, the walk would become so constrained that the walker will be trapped at the origin [see Fig.~1(b) for example]. The interesting questions are the existence of phase transition to the trapped trajectory at finite $\alpha_c$ and how the normal diffusive behavior at $\alpha=0$ and trapping at $\alpha=\alpha_c$ are interpolated as $\alpha$ increases. Thus in the following we focus on positive $\alpha$ regime.

To answer these questions, we perform extensive numerical simulations. To focus on the role of memory parameter, we fix $p=0$ and vary $\alpha$ in the range from $0$ to $1$ (Fig.~2). We measured the average dispersion of the walker in terms of the mean square displacement, $\langle r^2(t) \rangle$, as a function
of the number of steps $t$. $\langle \cdots\rangle$ denotes the ensemble average. The dispersion increases with number of steps in a power-law manner at long times,
\begin{equation}
\langle r^2(t)\rangle \sim t^{2\nu}~.
\end{equation}
The exponent $\nu$ is $1/2$ for ordinary diffusion, and $\nu<1/2$ indicates sub-diffusive behavior. We found that in one dimension, $\nu$ decreases continuously from $1/2$ at $\alpha=0$ to zero near $\alpha_c\approx0.5$ [Fig.~2(a)]. This suggests a trapping transition at finite $\alpha_c\approx0.5$ and sub-diffusion with continuously varying exponent $\nu$ for $0<\alpha<\alpha_c$.

\begin{figure}
\centering
\includegraphics[width=.9\linewidth]{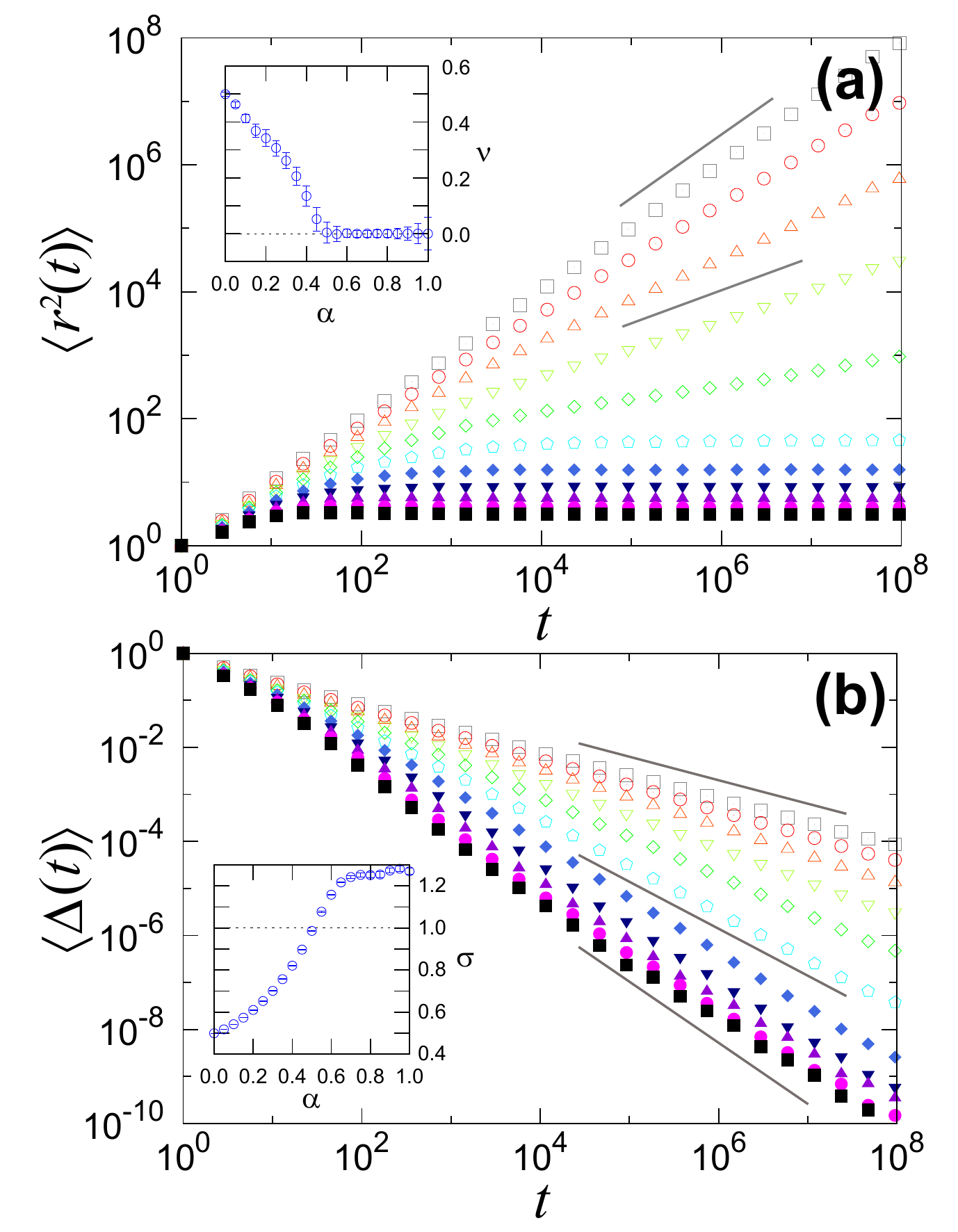}
\caption{Numerical simulation results for $0\le\alpha\le1$ and $p=0$ in one dimension.
(a) Mean square displacement $\langle r^2(t)\rangle$ as a function of  number of steps $t$ are displayed in double-logarithmic scale. Straight lines are guidelines with slopes $1$ (top) and $0.5$ (bottom). Different symbols correspond to different $\alpha$ from $0$ (top-most) to $1$ (bottom-most) in the interval of $0.1$, from top to bottom. (Inset) The exponent $\nu$ vs.\ the memory parameter $\alpha$, obtained by fitting the data to Eq.~(2) for large $t$. 
(b) Double-logarithmic plot of the probability of visiting an unvisited site at step $t$, $\langle\Delta(t)\rangle$, as a function of $t$. Straight lines are guidelines with slopes $-0.5$, $-1$, and $-1.3$, from top to bottom. Same symbols as in (a) are used.
(Inset) The exponent $\sigma$ vs.\ $\alpha$, obtained by fitting the data to Eq.~(3) for large $t$. Ensemble averages were taken over $10^4$ independent trajectories.
}
\end{figure}

We also measured the average probability that the walker visits to an unvisited site at step $t$, $\langle\Delta(t)\rangle$.
For one dimensional simple symmetric random walk ($\alpha=0$), it is known  \cite{hughes} that $\langle\Delta(t)\rangle$ decreases with $t$ as a power law,
\begin{equation}
\langle\Delta(t)\rangle \sim t^{-\sigma}~,
\end{equation}
with $\sigma=1/2$.
We found that $\sigma$ increases with $\alpha$, 
becomes $\sigma=1$ at $\alpha=\alpha_c\approx0.5$, and further increases for $\alpha>\alpha_c$ [Fig. 2(b)].
This indicates that the average total number of visited sites $\langle S(t)\rangle$ upto $t$-steps, given by the sum of $\langle\Delta(t)\rangle$ over $t$, exhibits distinct behaviors depending on $\alpha$.
$\langle S(t)\rangle$ diverges as $\sim t^{1-\sigma(\alpha)}$ for $\alpha<\alpha_c$ (diffusing), whereas it becomes finite for $\alpha>\alpha_c$ (trapped).
At the critical point $\alpha=\alpha_c$, $\langle S(t)\rangle$ increases logarithmically. This result is also in support of the existence of a trapping transition at finite $\alpha_c$. 

Thus, with the memory parameter we can obtain the slow or limited mobility in walker's motion. Depending on its strength, it can kinetically induce anomalous sub-diffusion or trapping, and transition between them. The heterogeneity of individual trajectory would follow at the critical point. It can also be thought that the memory parameter $\alpha$ can take different values person to person, as it characterizes an individual's degree of revisitation tendency. However, with the memory parameter alone, the walker is either diffusing or gets trapped at some localized region in space. Ultraslow dispersion can only be expected for a single specfic value of the memory parameter at the critical point. 

\begin{figure}
\centering
\includegraphics[width=.9\linewidth]{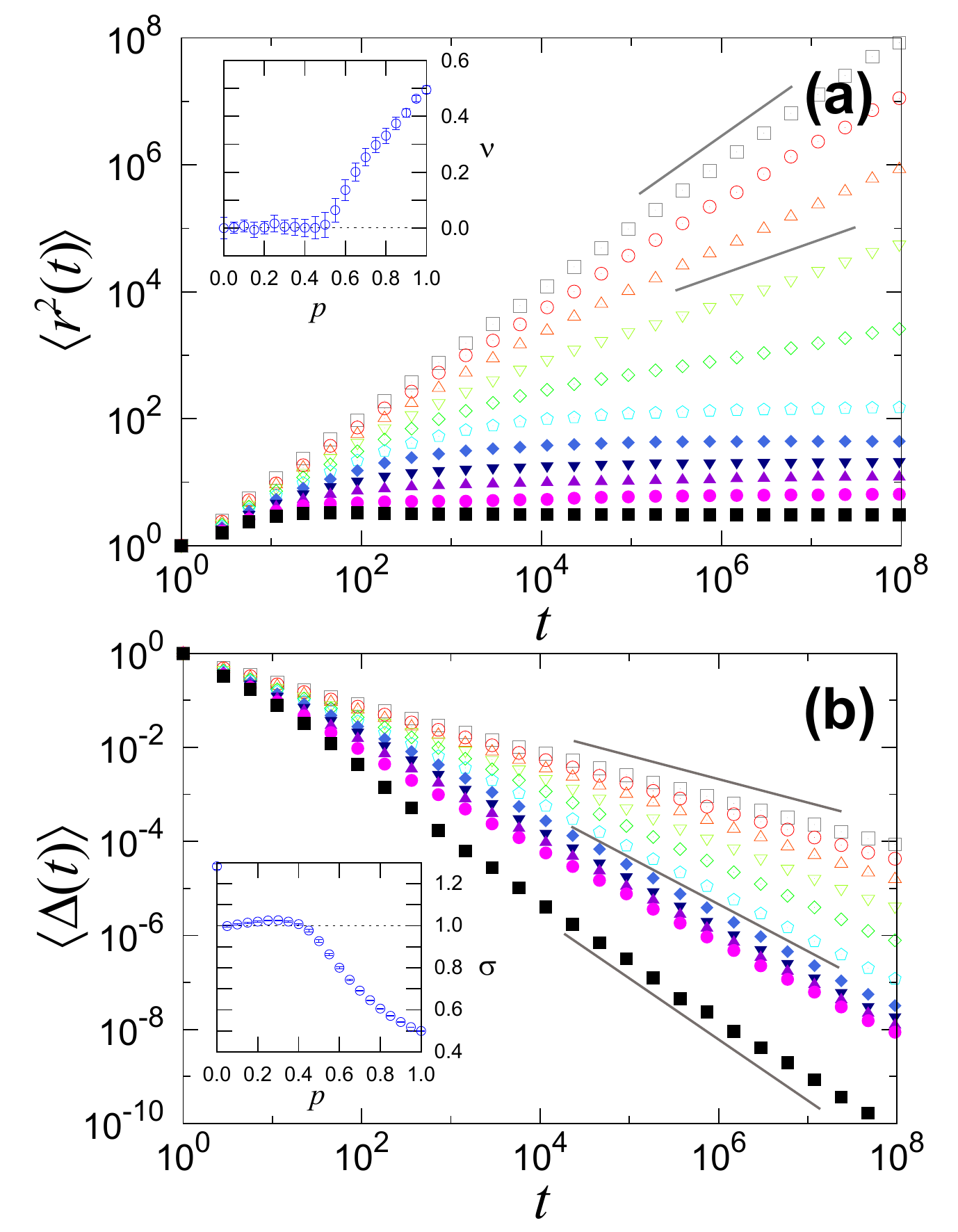}
\caption{Numerical simulation results for $\alpha=1$ and $0\le p\le1$ in one dimension.
(a) $\langle r^2(t)\rangle$ vs.\ t in double-logarithmic scale. Straight lines are guidelines with slopes $1$ (top) and $0.5$ (bottom). Different symbols correspond to different value of $p$ from $1$ (top-most) to $0$ (bottom-most) with the interval of $0.1$, from top to bottom. (Inset) The exponent $\nu$ vs.\ $\alpha$, obtained by fitting the data to Eq.~(2) for large $t$. 
(b) Double-logarithmic plot of $\langle\Delta(t)\rangle$ vs.\ $t$. Straight lines are guidelines with slopes $0.5$, $1$, and $1.3$, from top to bottom. Same symbols as in (a) are used.
(Inset) The exponent $\sigma$ vs.\ $\alpha$, from fitting the data to Eq.~(3) for large $t$. }
\end{figure}

\section{Role of impulse parameter}

The impulse parameter $p$ models the degree of impulsive, random decisions in walker's motion, introducing occasional unbiased random jumps.
Intuitively, introduction of unbiased random jumps $(p>0)$ will give rise to larger dispersion of the walker's trajectory than $p=0$ case. A question of interest is what the effect of random jumps is to a trapped walker. To answer this question, we fix the memory parameter to be $\alpha=1$, high enough for a strong trapping, as well as in accordance with the empirical data \cite{song2}, and vary the impulse parameter $p$ in the range from $0$ to $1$ (Fig.~3).

By measuring the mean square displacement $\langle r^2(t)\rangle$ as a function of $t$, we find that there exists a critical impulse parameter $p_c\approx0.5$, above which the walker is no longer confined in the trap and diffuses [Fig.~3(a)]. For $p>p_c$, $\nu$ increases continuously from zero and becomes $1/2$ for $p\to1$, crossing-over to normal diffusion. This result is corroborated by the average probability of visiting an unvisited site $\langle \Delta(t)\rangle$ decaying as Eq.~(3) with the exponent $\sigma$ varying from $1$ to $1/2$ as $p$ increases from $p_c\approx0.5$ to $1$ [Fig.~3(b)]. 

The role of impulse parameter for $p<p_c$ is more intriguing.
Meausuring $\nu$, we have $\nu\approx0$ for $p<p_c$, suggesting that the walker is still trapped [Fig.~3(a)]. However, $\sigma$ is stuck at $\sigma=1$, that is, $\langle \Delta(t)\rangle\sim t^{-1}$, regardless of $p<p_c$ [Fig.~3(b)]. This means that the average total number of visited sites $\langle S(t)\rangle$ still increases logarithmically, $\langle S(t)\rangle\sim \log t$, suggesting that the walker is in fact marginally escaping from the trap.
Such behavior can be seen from the example trajectory shown in Fig.~1(c), where one can identify more than one traps. 

\begin{figure}
\includegraphics[width=\linewidth]{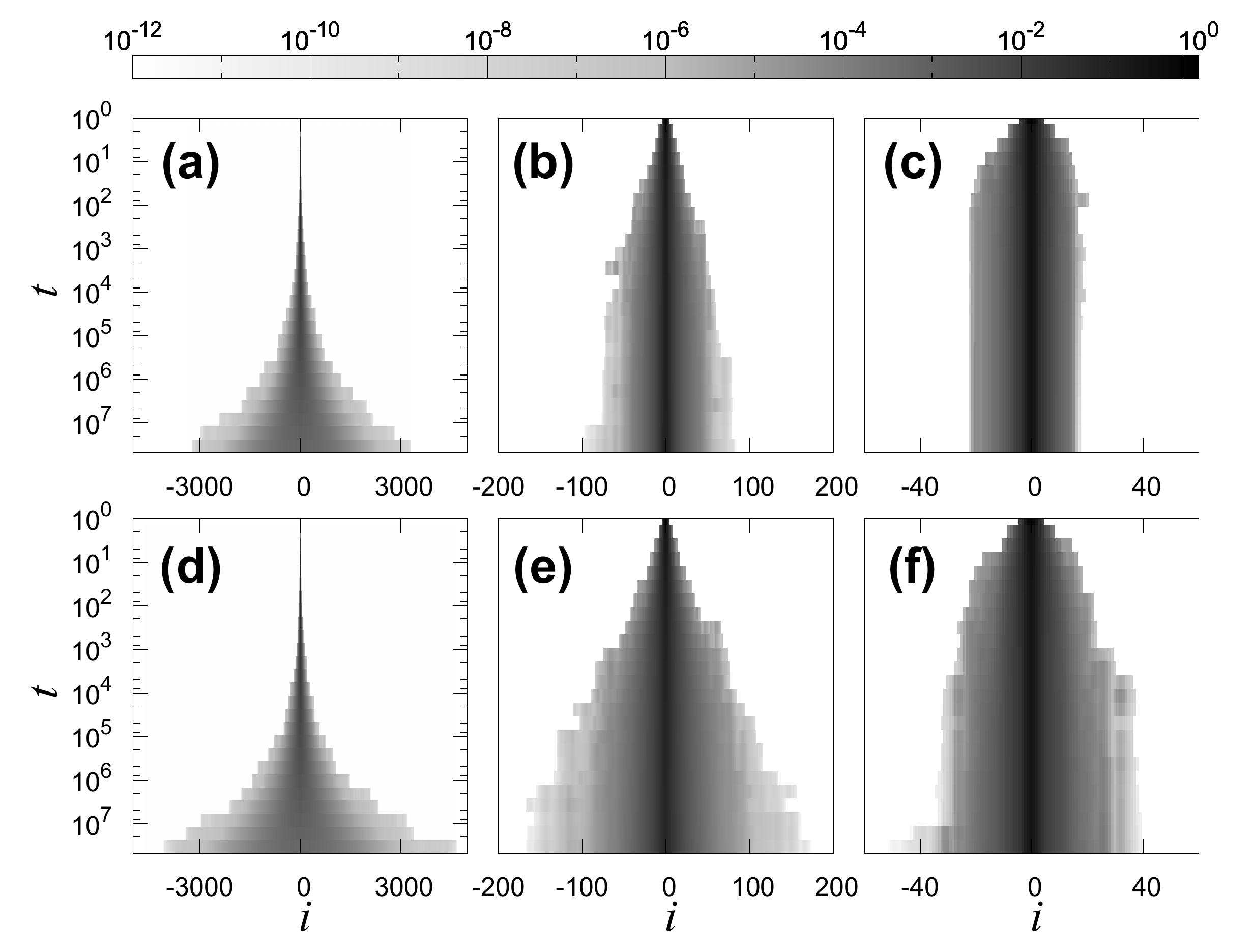}
\caption{Color-coded plot of the time evolution of the probability distribution $P_i(t)$ that the walker will be at site $i$ at step $t$, for parameter combinations $(\alpha,p)$ as (a) $(0.2,0)$, (b) $(0.5,0)$, (c) $(0.8,0)$, (d) $(0.8,1)$, (e) $(0.5,1)$, and (f) $(0.2,1)$, representative of the sub-diffusive (a, d), critical (b, e), trapped (c), and logarithmc diffusive (f) regimes. Note the logarithmic scale in time directions.
}
\end{figure}

To illustrate mobility patterns exhibited by different phases, we calculated the probability distribution $P_i(t)$, defined as the average probability to find the walker starting from origin at site $i$ after $t$-steps (Fig.~4). For the memory-only model $(p=0)$, we have diffusing ($\alpha<\alpha_c$), critical $(\alpha=\alpha_c)$, and trapped phases $(\alpha>\alpha_c)$ [Figs.~4(a--c), respectively]. In particular, in the trapped phase, $P_i(t)$ becomes time-independent at long times, suggesting the walker is competely trapped, visiting a set of trapping sites repeatedly {\it ad infinitum}. With the impulse parameter, the behavior is similar for the diffusing phase [Fig.~4(d)]. For $p\le p_c$, the random jumps by the impulse parameter leads to larger dispersion in $P_i(t)$ [Figs.~4(e,f)]. In particular, $P_i(t)$ is not strictly stationary for $0<p<p_c$ but gets dispersed with time very slowly (in logarithmic orders) [Fig.~4(f)].

Such logarithmic dispersion is reminiscent of the Sinai diffusion in quenched-disordered media \cite{sinai}. The revisitation tendency kinetically generates random traps, at which the walker is trapped in short timescales; thanks to random jumps, however, the walker can slowly escape from the trap, the timescale of which is of exponential order in the trap strength \cite{sinai}, but the walker will soon get re-trapped at another kinetically-generated trap. Such a series of trapping leads to the observed ultraslow dispersion. In our model the trap strength changes dynamically, making it even slower than $\sim (\log t)^4$ dispersion for the original Sinai diffusion with static disorder \cite{sinai}.

\section{Kinetic origins of the anomalous diffusion}
Our numerical simulation results show that the memory parameter $\alpha$ can modulate the diffusivity of the mobility with memory, from sub-diffusion to trapping. In order to gain more microscopic understanding how such limited mobility is generated by the memory, we measured the directional correlation function, $C(s)$, defined as the average correlation between the directions of two steps separated by the lag of $s$ steps, that is, $C(s)=\langle\overline{\eta(t)\eta(t+s)}\rangle$, where $\eta(t)=r(t)-r(t-1)$ is the step direction in one dimension and the overbar denotes the time average. In simple symmetric random walk $(\alpha=0)$, $C(s)=0$. We found that for $\alpha>0$, this directional correlation is non-zero. 

In the sub-diffusive regime ($0<\alpha\lesssim0.5$), two notable features emerge for $C(s)$ [Fig.~5(a)]. First, $C(s)$ is on average negative in the entire range of $s$, and its absolute value for small $s$ increases with $\alpha$. This means that the walk becomes more and more strongly anti-biased towards the current position anytime. Second, $|C(s)|$ decays as a power law, $|C(s)|\sim s^{-\zeta}$, with a crossover between two regimes. The power-law exponent is $\zeta\approx1/2$ for small $s$ and $\zeta\approx2$ for large $s$, and the crossover timescale decreases with $\alpha$. This long-ranged directional anti-correlations might be the microscopic origin for  the anomalous sub-diffusive behavior, with a similar mechanism as in the continuous-time random walks with anti-correlated fractional Gaussian noise \cite{barkai}. Meanwhile, in the trapped regime ($\alpha\gtrsim 0.5$), the directional correlation develops an alternating behavior around zero, reflecting the back-and-forth motion within the trap [Fig.~5(a), inset]. The amplitude of alternation increases with $\alpha$, as the trapping becomes stronger. 

\begin{figure}
\centering
\includegraphics[width=.9\linewidth]{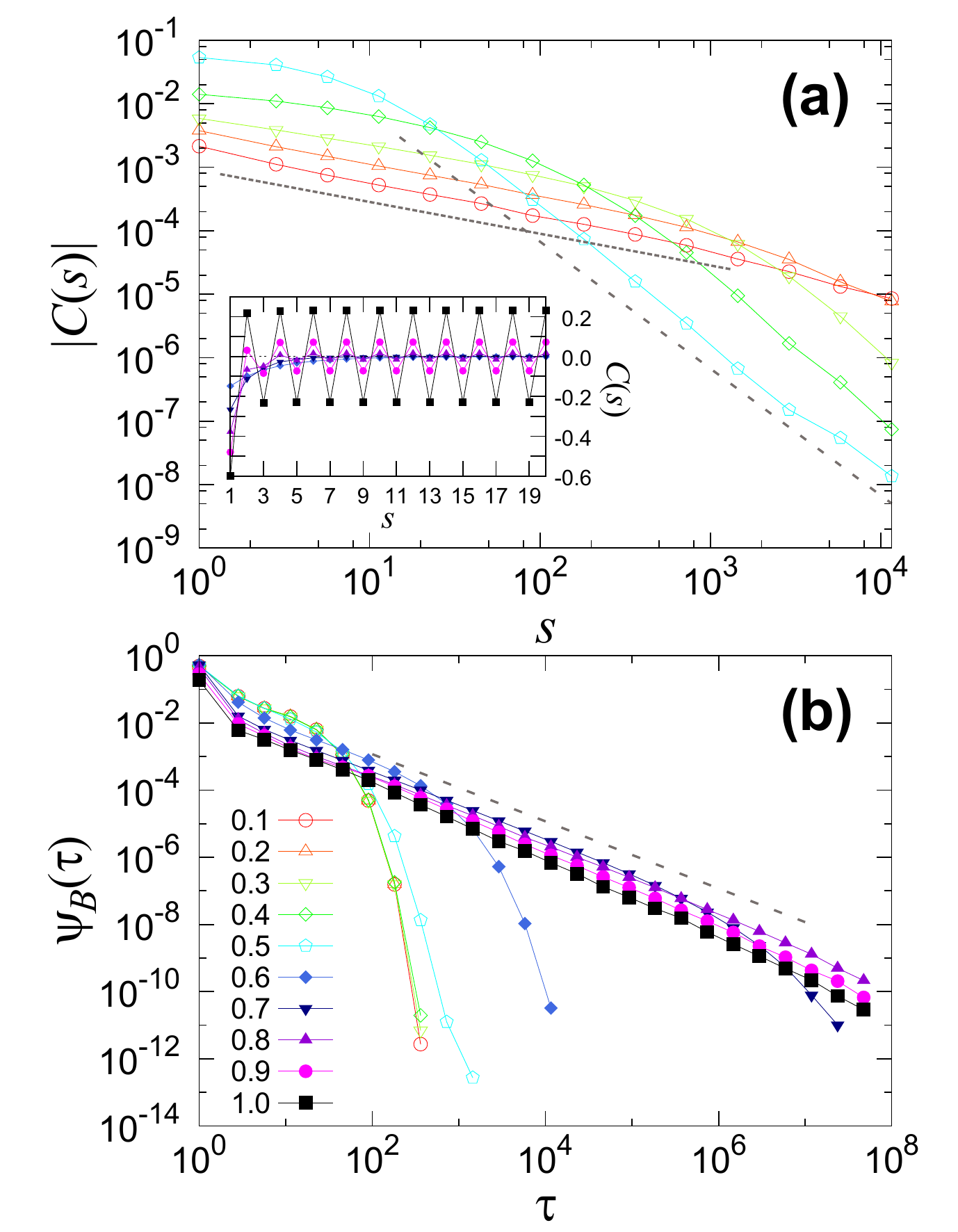}
\caption{Detailed kinetics of the model. (a) Directional correlation function $C(s)$ as a function of the step lag $s$ in double-logarithmic plot. Straight lines are guidelines with slopes $-0.5$ (dotted) and $-2$ (dashed). (b) Block waiting time distribution $\psi_B(\tau)$ with the blocksize $\ell=7$ in double-logarithmic scale. Straight line is a guideline with slopes $-1$. Same set of symbols are used in (a) and (b).}
\end{figure}

A well-known mechanism for sub-diffusion is the power-law distribution of waiting times between consecutive steps, or pausing times \cite{montroll}.
When the unbiased random walk with finite step length has the waiting time distribution $\psi(\tau)$ of a power-law form, $\psi (\tau) \sim \tau^{-(1+\mu)}$, with $0<\mu<1$, its dispersion becomes sub-diffusive as
$\langle r^2 \rangle \sim t^{\mu}$ at long times. To consider how such an effect manifests in our model, we divide the lattice into non-overlapping linear blocks of size $\ell$ (typically $\ell=7$, to cover the typical size of a trap), and calculated the histogram of the time interval $\tau$ between two consecutive crossings of the block boundaries by the walker, which we call the block waiting time distribution $\psi_B(\tau)$.

We found that in the trapped regime ($\alpha\gtrsim 0.5$), the block waiting time distribution follows a power law, $\psi_B(\tau)\sim\tau^{-1}$, for large $\tau$  [Fig.~5(b)], leading diverging average waiting time. This divergence is likely rooted from the existence of strong trapping, with diverging mean escaping time. The power-law behavior with $\mu=0$ in the trapped regime is apparently consistent with the lack of dispersion $\nu=0$ in that regime. In the sub-diffusive regime ($0<\alpha\lesssim0.5$), however, the block waiting time distribution is found to decay exponentially, rather than as a power law [Fig.~5(b)].
This result shows that the sub-diffusive mobility of the walker in our model is quite different from that of an unbiased random walker with power-law waiting time distribution; our walker is not waiting at particular locations but moves more actively from block to block constantly. It is the anti-correlated motion revealed above that provides key kinetic mechanism behind the limited mobility.

\section{In two-dimensions}
We performed numerical simulations of our model in one dimension for other combinations of $(\alpha,p)$-parameters. We found that the critical value of memory parameter $\alpha_c$ depends on the impulse parameter $p$; starting from $\alpha_c\approx0.5$ at $p=0$, $\alpha_c$ increases with $p$ [Fig.~6(a)]. We also performed numerical simulations in two dimensional square lattice, finding overall similar behaviors. A notable difference is that the logarithmic dispersion for $\alpha>\alpha_c$ and $p>0$ is seen more prominent in two dimensions [Fig.~6(b)]. Our numerical data suggest the dispersion in the logarithmic regimes as $\sim (\log t)^{\gamma}$ with $\gamma\approx0.3$ in one dimension and $\gamma\approx1.5$ in two dimensions. This result shows that the effect of random jumps by the impulse parameter becomes manifestly more significant in higher dimensions. 

\begin{figure}
\includegraphics[width=\linewidth]{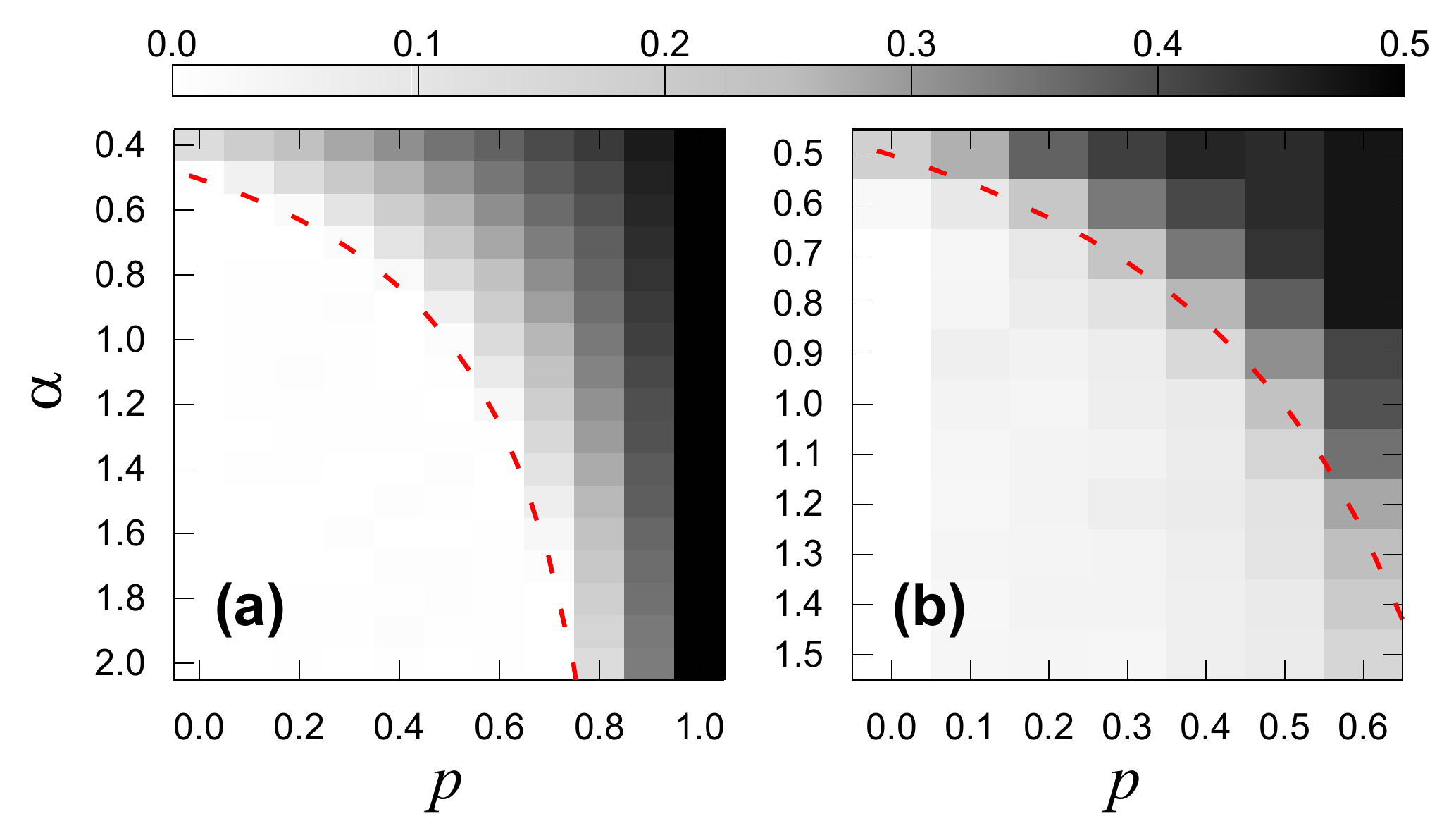}
\caption{Color-coded plots of the exponent $\nu$ in $(\alpha,p)$-parameter space in (a) one and (b) two dimensions. Dashed lines are guidelines of $\alpha(1-p)=1/2$.
}
\end{figure}

\section{Continuum limit}
The discrete-time master equation for $P_i(t)$ can be written as,
\begin{equation}
P_i(t+1)-P_i(t)={\sum_j}^{\prime} \left[ w_{i,j}(t)P_j(t)-w_{j,i}(t)P_i(t) \right],
\end{equation}
where the transition probability $w_{j,i}(t)$ is given by Eq.~(1) and the restricted summation runs over nearest neighbors $j$ of $i$.
Decomposing $w_{j,i}$ into memory-dependent and random jumps, we can rewrite the above equation as
\begin{align}
P_i(t+1)-P_i(t)&=  (1-p) {\sum_j}^{\prime}\left[ \mu_{i,j}(t)P_j(t)-\mu_{j,i}(t)P_i(t) \right] \nonumber\\
&\quad+p {\sum_j}^{\prime}\frac{1}{2d}\left[P_j(t)-P_i(t) \right].
\end{align}
Here $\mu_{j,i}(t)=\phi_j(t)/{\sum_k}^{\prime}\phi_k(t)$ with $\phi_j(t)=[n_j(t)+1]^{\alpha}$, denotes the memory-dependent jumping probability. Let us consider a continuous time and space limit of Eq.~(5). One can rewrite $\mu_{j,i}(t)$ in one dimension as $\mu_{i+1,i}(t)=\frac{1}{2}-\frac{1}{2}\frac{\phi_{i+1}(t)-\phi_{i-1}(t)}{\phi_{i+1}(t)+\phi_{i-1}(t)}\approx \frac{1}{2}-\frac{1}{2}\frac{\partial\phi(x,t)}{\partial x}/\phi(x,t)$, {\it etc.}, assuming that $\phi(x,t)$, the continuum analog of $\phi_i(t)$, is analytic at $x$.
This leads to the continuum limit of Eq.~(5) in one dimension as
\begin{equation}
\frac{\partial P(x,t)}{\partial t}
= \frac{1}{2}\frac{\partial ^2P(x,t) }{\partial x^2}
- (1-p) \frac{\partial}{\partial x} 
\left[ P(x,t) \frac{\partial \phi(x,t)/\partial x}{\phi(x,t)} \right].
\end{equation}
Taking advantage of the relation between $\phi(x,t)$ and $P(x,t)$, one finally obtains a highly-nonlinear integro-differential equation for $P(x,t)$,
\begin{align}
&\frac{\partial  P(x,t)}{\partial t} 
=\frac{1}{2}\frac{\partial ^2 P(x,t) }{\partial x^2}\nonumber\\
&\quad -\alpha(1-p) \frac{\partial}{\partial x} 
\left\{ P(x,t) \frac{\partial}{\partial x}\log\left[1+\int_0^t P(x,t')dt' \right] \right\}. 
\end{align}

Eq.~(7) is too nonlinear to be solved exactly, yet it can be used to locate the transition point between diffusion and trapping as follows. Approaching to the transition point, $P(x,t)$ starts to become time-independent. Looking for the condition of time-independent $P(x,t)$ for large $t$, equivalently $n(x,t)\approx tP(x)$, we find the transition point to trapping as $\alpha_c(1-p)=1/2$, which is in good agreement with the numerical simulation results (Fig.~6).

\section{Summary and discussion}
We have studied a random walk model in which the jumping probability to a site is dependent on the number of previous visits to the site, as a model of the mobility with memory, such as that of a human individual or an animal. Through extensive numerical simulations, we found that by varying two parameters called the memory parameter $\alpha$ and the impulse parameter $p$, various limited mobility patterns such as the sub-diffusion, trapping, and logarithmic diffusion could be observed.  
With the memory parameter $\alpha$, a long-ranged directional anti-correlation is developed in the walker's motion, resulting in anomalous sub-diffusive behaviors. At the critical value of $\alpha_c$, the transition from sub-diffusive to trapped motion occurs, above which the walker's trajectory becomes confined in a localized trapping region. By allowing random jumps with the impulse parameter $p$, the walker can escape from the trap, albeit very slowly, but gets trapped to another trap kinetically-generated by the walker's own motion, resulting in an ultraslow logarithmic diffusive behavior.

Our results demonstrate that the memory of walker's has-beens can be a mechanism explaining many of empirical characteristic patterns of human mobility \cite{brockmann,gonzalez,song}. The revisitation tendency driven by the memory has also been proposed and supported by empirical data in the more detailed, data-driven modeling of human mobility \cite{song2}. Our work extends this approach to encompass broader spectrum of limited mobility patterns, including sub-diffusion and trapping, and the transition between them. It would be interesting to see if our modeling framework with the memory and impulse parameters could be applied to various empirical data for human and animal's mobility, which might uncover further universality and specificity in the physics of mobility of animated objects \cite{foraging}.

\acknowledgments
This work was supported by Basic Science Research Program through NRF
grant funded by MEST (No.\ 2011-0014191).

\end{document}